\documentclass[aps,prd,floatfix,preprint,nofootinbib]{revtex4-1}
\usepackage{graphicx}
\usepackage{epic}
\usepackage{eepic}
\usepackage{latexsym}
\usepackage{amssymb,amsmath}

\newcommand{\eq}[1]{(\ref{#1})}
\newcommand{\be}{\begin{equation}}
\newcommand{\ee}{\end{equation}}
\newcommand{\bea}{\begin{eqnarray}}
\newcommand{\eea}{\end{eqnarray}}

\newcommand{\hs}[1]{\hspace{#1 mm}}

\newcommand{\vc}{\vec{k}}
\newcommand{\vx}{\vec{x}}
\newcommand{\vv}{|0\right>}
\newcommand{\vvc}{\left<0|}
\newcommand{\tk}{\tilde{k}}
\newcommand{\hk}{\hat{k}}

\def\a{\alpha}
\def\b{\beta}

\def\C{\Gamma}
\def\d{\delta}

\def\e{\epsilon}

\def\f{\phi}
\def\fr{\frac}

\def\L{\Lambda}
\def\m{\mu}
\def\n{\nu}

\def\r{\rho}

\def\th{\theta}

\def\O{\Omega}

\def\del{\partial}

\let\bm=\bibitem
\def\nn{\nonumber}

\begin{document}

\title{Superhorizon Electromagnetic Field Background from Higgs Loops in Inflation} 

\author{Ali Kaya}
\email[]{ali.kaya@boun.edu.tr}
\affiliation{Bo\~{g}azi\c{c}i University, Department of Physics, 34342, Bebek, \.{I}stanbul, Turkey}

\date{\today}

\begin{abstract}

If Higgs is a spectator scalar, i.e. if it is not directly coupled to the inflaton, superhorizon Higgs modes must have been exited during inflation. Since Higgs is unstable its decay into photons is expected to seed superhorizon photon modes. We use in-in perturbation theory to show that this naive physical expectation is indeed fulfilled via loop effects. Specifically, we calculate the first order Higgs loop correction to the magnetic field power spectrum evaluated at some late time after inflation. It turns out that this loop correction becomes much larger than the tree-level power spectrum at the superhorizon scales. This suggests a mechanism to generate cosmologically interesting superhorizon vector modes by scalar-vector interactions. 

\end{abstract}

\maketitle

\section{Introduction}

Inflation is the leading theory of the early universe genuinely yielding scale free cosmological perturbations. As there are also alternative viable theories, it is desirable to find distinguishing features unique to inflation to single out the correct early universe paradigm. The main aim of this work is to suggest one such feature based on the following simple observation: If Higgs is a spectator not directly coupled to the inflaton, it would behave like a (nearly) massless scalar and together with cosmological perturbations superhorizon Higgs modes are also created during inflation . One then naturally anticipates superhorizon photon modes to be exited due to Higgs-photon interactions. In this paper we show that this naive physical expectation is indeed realized by loop effects. 

Since the electromagnetic field is conformally coupled to gravity, its own power spectrum is the same with the flat space one and no cosmologically interesting behavior may arise. Nevertheless, we find that the Higgs modes running in the loops of the photon two point function greatly enhances the power spectrum at the superhorizon scales, which may produce potentially interesting superhorizon physics for the photons after inflation. There are many works studying  quantum dynamics of the electromagnetic field coupled to the scalars both in the pure de Sitter space and slow-roll inflationary backgrounds. These works mainly focus on issues arising during inflation like  photon mass generation (see e.g. \cite{w1,w2,w3,w4,w5,w6}; see also \cite{s0,s1,s2,s3,s4,s5,s6,s7,s8,s9} for studies related to the scalar field dynamics in an expanding universe). Our aim here is to pin down a loop effect extending beyond inflation to some later epoch of interest. One may think that this requires  a detailed and presumably analytic information about the mode functions, which evolve in a complicated way after inflation. This is not a problem for the electromagnetic field since the photons are conformally coupled to gravity and the photon mode functions are the same with the flat space counterparts throughout all cosmological evolution. The main observation of this paper is that the late time Higgs loop correction can be divided into the sum of different terms (roughly) corresponding to the individual epochs and the one associated with the inflation is the largest. As a result, it is possible to obtain the main loop contribution after inflation by using the well known inflationary scalar mode functions without the need of extending them beyond. 

Like any loop effect infinities also arise here, which must be suitably regularized for proper interpretation. In this paper we apply adiabatic regularization \cite{ad1,ad2} that works by subtracting certain divergent terms determined by an adiabaticity condition related to the expansion rate of the universe where the zeroth order contribution is the flat space one. Fortunately, the massless scalar mode function in de Sitter space, which  appears in our loop computation,  has clearly identifiable adiabatic zeroth order flat space and adiabatic first order expanding pieces. This decomposition makes the adiabatic subtraction scheme easy to utilize in our problem. In the appendix we also demonstrate that dimensional regularization concurs qualitatively with the adiabatic method, which shows the robustness of our findings. 

\section{The Model and the Loop Corrections}

Higgs physics is rich in flat space and even a richer phenomenon is anticipated to emerge in cosmology. For example, the problem of (global or local) symmetry breaking in an expanding universe is not yet fully understood since it involves quantum corrections to the scalar potential and the scalar vacuum expectation value, both are nontrivially modified by the expansion of the universe. In this paper, we are only interested in Higgs's first order impact on the otherwise free electromagnetic field, and thus complicated issues about Higgs physics do not concern us. Moreover, since we will only use the inflationary scalar field mode functions, the mass of the Higgs is ignorable in our computations even though it may not be negligible compared to the Hubble scale of the late time of interest.  

At low energies after gauge symmetry breaking, the charged Higgs doublet was eaten by $W^\pm$ and $Z$ bosons as longitudinal components leaving only a real neutral scalar not directly interacting with photons. However, the presumed scale of inflation is much larger than the  symmetry breaking (electroweak) scale and one can safely neglect the Higgs vacuum expectation value and consider the unbroken theory. 

To sum up we work with the following Lagrangian
\be
L=-\fr14\sqrt{-g}F^{\m\n}F_{\m\n}-\sqrt{-g}(D_\m\f) (D^\m\f)^*,\label{1}
\ee
where $\f$ is the complex scalar representing the Higgs field, $D_\m\f=\del_\m\f-ieA_\m\f$ and $e$ is the dimensionless Higgs-photon coupling constant (in the unbroken theory). These fields are assumed to propagate on a fixed cosmological background having the metric
\be
ds^2=-dt^2+a(t)^2dx^idx^i,
\ee
where in a realistic scenario the scale factor $a(t)$ must describe the epochs like slow-roll inflation, reheating, radiation, matter and the recent accelerated expansion phases. The conformal time is defined as usual by
\be
d\eta=\fr{dt}{a}.\label{2}
\ee
Below, we will view the conformal time $\eta$ as a function of the proper time $t$. Because only the difference of two conformal times will appear in our results, there is no need to fix the undetermined integration constant in expressing $\eta$ in terms of $t$ from \eq{2}. 

In our calculation, we prefer to impose the Coulomb gauge
\be
\del_iA_i=0,\label{3}
\ee
which is suitable for canonical quantization. In the Coulomb gauge $A_0$ becomes a nonlocal composite field that can (perturbatively) be solved from its own equations of motion that reads
\be
\left(\del^2-2e^2a^2\f^*\f\right)A_0=iea^2\left(\f^*\dot{\f}-\f\dot{\f}^*\right),\label{4}
\ee
where the dot denotes the time derivative. Note that $A_0$ is totally fixed by the scalar field $A_0=A_0(\f)$ and plugging this solution back in the Lagrangian only generates (nonlocal) $\f$-self interactions. The canonical quantization of the remaining fields can be achieved by the mode expansions
\bea
&&A_i=(2\pi)^{-3/2}\int d^3k\,e^{i\vec{k}.\vec{x}-ik\eta}\,a^s_{\vc}\,\e^s_i(\vc)+h.c.\hs{15}s=1,2\nn\\
&&\f=(2\pi)^{-3/2}\int d^3k\left[e^{i\vc.\vx}\,\f_k(t)\,a_{\vc}+e^{-i\vc.\vx}\,\f_k^*(t)\,b^\dagger_{\vc}\right],\label{5}
\eea
where the mode function $\f_k$ and the polarization tensor $\e_i^s$ obey 
\bea
&&k^i\e_i^s=0,\hs{10}\e_i^s\e_j^{s*}=\fr{1}{2k}\left(\d_{ij}-\fr{k_ik_j}{k^2}\right),\nn\\
&&\ddot{\f}_k+3\,\fr{\dot{a}}{a}\,\dot{\f}_k+\fr{k^2}{a^2}\f_k=0,\hs{10}\f_k\dot{\f}_k^*-\f_k^*\dot{\f}_k=\fr{i}{a^3}.\label{6}
\eea
One may observe that while the photon mode function is completely determined in \eq{5} since $A_i$  obeys the free field equation in conformal time  $A_i^{''}-\del^2A_i=0$, the scalar mode function $\f_k$ needs to be specified from \eq{6}.  As usual the vacuum is defined by
\be
a_{\vc}^s\left. \vv =a_{\vc}\left. \vv=b_{\vc}\left.\vv=0,
\ee
which is connected to the selection of the scalar field mode function. The interaction Hamiltonian for the Higgs-photon coupling can be found from \eq{1} as
\be
H_I=e^2\int d^3 x\,a \,A_iA_i\f^*\f+ie\,\int\,d^3x \,a \,A_i\left(\f^*\del_i\f-\f\del_i\f^*\right),\label{7}
\ee
where the fields are in the interaction picture given in \eq{5}. 

One may introduce the polarization tensor 
\be
\vvc A_i(\vx,t)A_j(\vec{y},t)\vv=(2\pi)^{-3}\int d^3 k\,e^{i\vc.(\vx-\vec{y})}\,\Pi_{ij}(\vc,t),\label{8}
\ee
obeying $k^i\Pi_{ij}=0$ (here, $A_i$ obviously denotes the full Heisenberg picture operator). In this paper we concentrate on the gauge invariant magnetic field power spectrum $P(k,t)$, which can be defined as
\be
\vvc B_i(\vx,t)B_i(\vec{y},t)\vv=(2\pi)^{-3}\int d^3 k\,e^{i\vc.(\vx-\vec{y})}\,P(k,t),\label{9}
\ee
where $P(k,t)=k^2\Pi_{ii}$. The function $P(k,t)$  roughly gives the variance of the quantum magnetic field fluctuations at the comoving scale $k$ at time $t$. Since the electric field involves $A_0$, which becomes a composite operator in the Coulomb gauge, we prefer to focus on the magnetic field. One would expect the electric and the magnetic field fluctuations to be similar and, for example, to contribute comparably to the electromagnetic field energy density.

In \eq{9}, we make dot product of two vectorial quantities that are defined at different tangent spaces in a curved manifold.  If $t_0$ denotes the time of interest (today or may be some other earlier time), we normalize 
\be
a(t_0)=1\label{10}
\ee
so that the vectors in \eq{9} can be thought to live in the flat space, at least in a local Hubble patch, and the dot product becomes well defined having the usual geometrical meaning. With the normalization \eq{10}, the energy density of the magnetic field fluctuations can be expressed in terms of the power spectrum as
\be
\r_B(t_0)=\fr12\vvc B_iB_i\vv=\fr{1}{4\pi^2}\int_0^\infty dk\, k^2\,P(k,t_0).\label{12}
\ee
Note that the comoving and the physical scales are identical  at time $t_0$. 

\begin{figure}
\centerline{
\includegraphics[width=5.5cm]{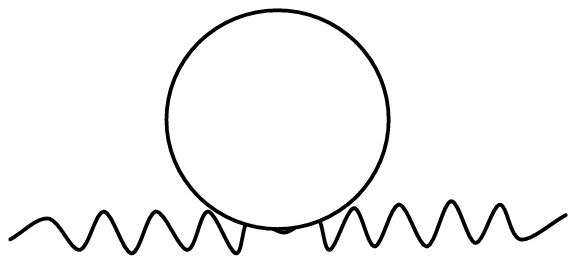}\hs{15}\includegraphics[width=6cm]{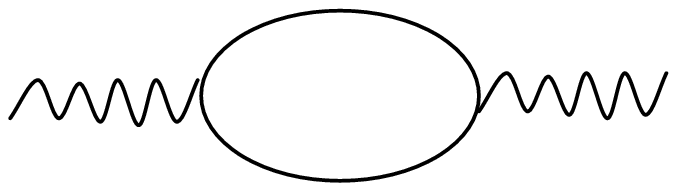}}
\caption{The diagrams corresponding to the one loop Higgs corrections to the photon two point function. The graphs on the left and on the right arise from the first and the second terms of the interaction Hamiltonian \eq{7}, respectively.}
\label{fig1}
\end{figure}

We are now ready to apply the in-in perturbation theory and we find it convenient to use Weinberg's commutator formula \cite{wein1}
\be\label{13}
O_H(t_0)=O_I(t_0)-i\int_{t_i}^{t_0} dt'\,[O_I(t_0),H_I(t')]-\int_{t_i}^{t_0} dt''\int_{t''}^{t_0} dt'\,[[O_I(t_0),H_I(t')],H_I(t'')]+. . . 
\ee
where $t_i$ is the initial time, $O_H$ is the Heisenberg picture field, $O_I$ is the corresponding free field in the interaction picture and $H_I$ is the interaction Hamiltonian. Eq. \eq{13} can be obtained by comparing the unitary time evolutions of $O_H$ and $O_I$. Using the interaction Hamiltonian \eq{7}, a straightforward but lengthy calculation gives the power spectrum in \eq{9} as
\bea
P(k,t_0)&&=k-e^2\int_{t_i}^{t_0}dt' a(t')\vvc \f^\dagger\f(t')\vv\sin[2k(\eta_0-\eta')]+\label{14}\\
&& 4ie^2\int_{t_i}^{t_0}dt''\int_{t''}^{t_0}dt'a(t')a(t'')\sin[k(\eta_0-\eta')]\left[F(k,t',t'')e^{-ik(\eta_0-\eta'')}-c.c.\right]+O\left(e^4\right),\nn
\eea
where $\eta_0$, $\eta'$ and $\eta''$ are conformal times corresponding to $t_0$, $t'$ and $t''$,   
\be
\vvc \f^\dagger\f(t')\vv=\fr{1}{2\pi^2}\int_0^\infty \,dk\, k^2|\f_k(t')|^2,\label{15}
\ee
and
\be
F(k,t',t'')=(2\pi)^{-3}\int d^3\tk\,\,\tk_i\,\tk_j\left(\d_{ij}-\fr{k_ik_j}{k^2}\right)\,\f_{\tk}(t')\,\f_{\tk}^*(t'')\,\f_{|\vec{k}+\vec{\tk}|}(t')\,\f_{|\vec{k}+\vec{\tk}|}^*(t'').\label{16}
\ee
The first term in \eq{14} is the tree-level contribution that is identical to the flat space one. The second and the third terms in \eq{14} come from the first and the second terms of $H_I$ in \eq{7}, respectively. These can be pictured as in Fig.\ref{fig1}.

The loop integrals in \eq{15} and \eq{16} contain UV (and possibly IR) divergences that must be regularized for proper physical interpretation, as we discuss in the next section. However, even after $P(k,t_0)$ is regularized and become finite, the field variance in position space  or the corresponding energy density may still be infinite. These are actually distributional divergences related to the coincidence limit and they show up even in the free theory. In cosmology, this problem is usually avoided by introducing a window function, which cuts out  the short scale UV modes. The window function is normalized so that
\be
\int d^3 x \,W(\vec{x})=1,\label{17}
\ee
and the smoothed out magnetic field can be defined as
\be
B_{Wi}(\vx)=\int d^3 y \,W(\vec{x}-\vec{y})\,B_i(\vec{y}).\label{18}
\ee
The corresponding variance becomes
\be
\vvc B_W^2\vv=\int d^3 k \,|{\cal W}(\vc)|^2\,P(k),\label{19}
\ee
where 
\be
{\cal W}(\vc)=(2\pi)^{-3/2}\int d^3 x \,e^{-i\vc.\vx}\,W(\vx).\label{20}
\ee
In cosmology one usually employs a Gaussian window function, but in this paper we prefer to choose 
\be
W(\vx)=\fr{8\e}{\pi^2(\e^2+4x^2)^2},\hs{15}{\cal W}(\vc)=(2\pi)^{-3/2}\,e^{-\e\, k/2},\label{21}
\ee
which exponentially cuts the momentum integrals at the scale $1/\e$. This choice will let us to obtain analytical expressions that at least in one case has finite $\e\to0$ limit giving a cutoff independent result. 

\section{Regularized Power Spectrum}

Eq. \eq{14} gives the magnetic field power spectrum on any cosmological background and for any specified scalar field mode function $\f_k$ obeying \eq{6}. Let us now focus on our main interest, a cosmology with inflation. For clarity, we only consider inflation and radiation epochs, which can be described by the following scale factor:
\be\label{22}
a(t)=\begin{cases}a_I\, e^{Ht}\hs{15}t_i\leq t\leq t_I \cr
\left(t/t_0\right)^{1/2}\hs{10}t_I\leq t\leq t_0,
\end{cases}
\ee
where $H$ is the Hubble scale of the inflation and $a_I$ is a constant. The inflation and the radiation eras correspond to the intervals $(t_i,t_I)$ and $(t_I,t_0)$, respectively. Note that $t_0$ can be any time of interest in the radiation epoch, it does not necessarily indicate the end of radiation. Matching the pieces of the scale factor at $t_I$ gives
\be
a_I=e^{-1/2}\sqrt{H_0/H},\hs{10}t_I=1/2H,\label{23}
\ee
where $H_0=1/2t_0$ is the Hubble parameter at $t_0$. 

Different mode functions $\f_k$ obeying \eq{6} define different vacuum states. The Bunch-Davies vacuum is physically the most relevant one and the corresponding inflationary mode function is given by\footnote{Since Higgs mass is many orders of magnitude smaller than the presumed inflationary Hubble scale, it is completely negligible during inflation.}  
\be
\f_k=\fr{1}{a\sqrt{2k}}\left[1+\fr{iaH}{k}\right]\exp\left(\fr{ik}{Ha}\right),\hs{10}t_i\leq t\leq t_I.\label{24}
\ee
Following inflation, the time evolution of $\f_k$ must be determined by solving the mode equation in the radiation era with the initial conditions supplied by \eq{24} at $t=t_I$. 

It is evident from \eq{14} that the time integrals from $t_i$ to $t_0$ can be calculated piece by piece, i.e. from $t_i$ to $t_I$ and from $t_I$ to $t_0$. Similarly, the second time integral from $t''$ to $t_0$ can be decomposed in the same way.  Just by dimensional reasoning the magnitude of each piece must be proportional to the corresponding Hubble scale, as a result the largest contribution must come when all the integration variables run in inflation. Consequently, to get the leading order contribution one may approximate
\bea
P(k,t_0)&&\simeq k-e^2\int_{t_i}^{t_I}dt' a(t')\vvc \f^\dagger\f(t')\vv\sin[2k(\eta_0-\eta')]+\label{25}\\
&& 4ie^2\int_{t_i}^{t_I}dt''\int_{t''}^{t_I}dt'a(t')a(t'')\sin[k(\eta_0-\eta')]\left[F(k,t',t'')e^{-ik(\eta_0-\eta'')}-c.c.\right].\nn
\eea
It is important to emphasize that in \eq{25} only the upper limits of the integrations are cut at $t_I$ as compared to \eq{14} and the time of interest $t_0$ is still encoded through the corresponding conformal time $\eta_0$, which is actually carried out by the photon mode functions. Essentially, \eq{25}  does not give a loop effect at inflation, but it is the leading order inflationary addition to a loop effect evaluated at some later time $t_0$.  

One may now use the mode function \eq{24} to determine the loop correction and not surprisingly this yields infinities. Specifically, both $\vvc\f^\dagger\f\vv$ and $F(k,t,t'')$, which are  given in \eq{15} and \eq{16} respectively, diverge for the de Sitter mode function. Let us try to cure these infinities by using adiabatic regularization. Expressed in terms of the conformal time, the first term (in the square brackets) of the mode function \eq{24} is identical to the flat space one and the second term is induced by the de Sitter expansion. In the language of adiabatic regularization, the first term has adiabatic order zero since it does not contain any time derivatives and the second term has adiabatic order one because it includes only a single time derivative carried by $H$.  Adiabatic regularization works by subtracting the divergent terms, which are grouped according to their adiabaticity. For example, after using \eq{24} in \eq{16} one obtains a bunch terms whose adiabatic orders range from zero to four. While the adiabatic order zero group contains only the first piece of the mode function \eq{24}, the adiabatic order four group consists of the second piece. Each group has the same degree of divergence or it gives a convergent loop integral, and for regularization the divergent groups must be thrown away. 

Applying the above procedure to regularize $\vvc \f^\dagger\f\vv$ in \eq{15} one sees that even the highest adiabatic order two term logarithmically diverges and must be subtracted giving $\vvc \f^\dagger\f\vv=0$. This is a typical situation encountered in adiabatic regularization of massless fields; for example the energy-momentum tensor of a massless scalar propagating on a cosmological background also vanishes after adiabatic subtractions. To overcome this problem one should start from a massive field, apply adiabatic regularization and carry out the zero mass limit carefully (indeed, only using this procedure the well known trace anomaly can be obtained in this formalism). For the massive scalar field, the adiabatic regularization yields
\bea
&&\vvc \f^\dagger\f\vv_{reg}=\fr{1}{2\pi^2}\int_0^\infty \,dk\, k^2|\f_k^{(m)}|^2\label{mad}\\
&&-\fr{1}{4\pi^2a^3}\int_0^\infty\, \fr{dk\,k^2}{(m^2+k^2/a^2)}\left[1+\fr{H^2}{(m^2+k^2/a^2)}+\fr{3m^2H^2}{4(m^2+k^2/a^2)^2}-\fr{5m^4H^2}{8(m^2+k^2/a^2)^3}\right],\nn 
\eea
where the massive mode function is given by
\be\label{mm}
\f_k^{(m)}=\sqrt{\fr{\pi}{4Ha^3}}\exp(i\pi\n/2)H^{(1)}_\n(k/aH).
\ee
Here, $H_\n^{(1)}$ is the Hankel function of the first kind and $\n=\sqrt{9/4-m^2/H^2}$. For nonzero $m$, \eq{mad} is well defined both as $k\to0$ and $k\to\infty$. Thus, one may introduce dimensionless variables $m=H\hat{m}$ and $k=aH\hat{k}$ so that 
\be
\vvc \f^\dagger\f\vv_{reg}=c_1H^2,\label{26a}
\ee
where the numerical constant $c_1$ is given by 
\bea
c_1=\int_0^\infty d\hat{k}&&\left[\fr{\hat{k}^2}{8\pi}|H_{\hat{\n}}^{(1)}(\hat{k})|^2-\right.\label{clad}\\
&&\left.\fr{\hat{k}^2}{4\pi^2\sqrt{\hat{m}^2+\hat{k}^2}}\left(1+
\fr{1}{(\hat{m}^2+\hat{k}^2)}+\fr{3\hat{m}^2}{4(\hat{m}^2+\hat{k}^2)^2}-\fr{5\hat{m}^4}{8(\hat{m}^2+\hat{k}^2)^3}
\right)\right],\nn
\eea
$\hat{\n}=\sqrt{9/4-\hat{m}^2}$. For the massless field, it is now possible to take $\hat{m}\to0$ limit\footnote{Strictly speaking, adiabatic regularization only cures the UV divergence. In taking $\hat{m}\to0$ limit one still needs a viable IR cutoff, which is a well known requirement for massless fields in de Sitter space.} in \eq{clad}. Another way of motivating \eq{26a} is to keep (not too dangerous) logarithmically divergent adiabatic order two term in \eq{15} and introduce UV /IR cutoffs $\L_{UV}$/$\L_{IR}$, which then gives \eq{26a} with $c_1=\ln(\L_{UV}/\L_{IR})/(4\pi^2)$. 

As for the function $F(k,t',t'')$ given in \eq{16}, we find that only the adiabatic order four term is finite. After throwing out others for regularization we obtain
\be
F_{reg}(k,t',t'')=\fr{H^4}{16\pi^2k}\int_0^\infty l\,dl\, \int_{-1}^{1}du\fr{(1-u^2)}{(1+l^2+2lu)^{3/2}}\exp\left[ik(\eta''-\eta')\left(l+\sqrt{1+l^2+2lu}\right)\right],
\label{26}
\ee 
where $l$ is the dimensionless loop integration variable defined by $\tk=l\,k $ as compared to \eq{16} and $u=\cos\th$ where $\th$ is the angle between the vectors $\vec{k}$ and $\vec{\tk}$. Technically, the scaling $\tk=l\,k$ is only allowed for $k\not=0$, but \eq{25} already contains enough suppression factors as $k\to0$ so no problem arises at $k=0$. 

It might be possible to solve the integrals in \eq{26} in terms of special functions for all $k$. In the following we will try to get an approximate expression when $k$ is superhorizon. We first observe that the oscillating integrand in \eq{26}, which has no stationary phase, is both suppressed by $1/l^2$ and oscillates more for $l\gg1$. Therefore, for any given $k$ the largest contribution to the $l$-integral in \eq{26} must come around $l=O(1)$. We also observe that because of the scale factors the time integrals in \eq{25} are maximized near the end of inflation. Moreover, at early times in inflation the conformal time gets larger and the phase in \eq{26} oscillates more, which diminishes the integral. Choose now $k$ to be superhorizon at the end of inflation $k\ll H a(t_I)$. This implies near the end of inflation that $k(\eta'-\eta'')\ll1$. As a result, the phase in \eq{26} can be neglected (since $l$ can be thought to be order one) and one may approximate
\be
F_{reg}(k,t',t'')\simeq c_2\fr{H^4}{16\pi^2k},\hs{15}k\ll H a(t_I)\label{27}
\ee
where $c_2=\int_0^\infty l\,dl\, \int_{-1}^{1}du(1-u^2)/(1+l^2+2lu)^{3/2}$. Inserting \eq{27} in \eq{25}, the regularized power spectrum becomes
\bea
P_{reg}(k,t_0)&&\simeq k-c_1e^2H^2\int_{t_i}^{t_I}dt' a(t')\sin[2k(\eta_0-\eta')]+\label{28}\\
&& c_2\fr{e^2H^4}{2\pi^2k}\int_{t_i}^{t_I}dt''\int_{t''}^{t_I}dt'\,a(t')\,a(t'')\sin[k(\eta_0-\eta')]\sin[k(\eta_0-\eta'')], \nn
\eea
which is valid when $k\ll H a(t_I)$.

To proceed, using the scale factor \eq{22} one may find that
\be
\eta_0-\eta'=\fr{1}{\sqrt{HH_0}}\left[e^{H(t_I-t')}-2\right]+\fr{1}{H_0},\label{29}
\ee
where $t'$, which has the corresponding conformal time $\eta'$, is some time during inflation. Therefore, at a late time interest satisfying $H_0\ll H$ one gets\footnote{Here, we assume that $t'$ does {\it not} refer to an early time in inflation since we use \eq{30} in \eq{28} where the time integrals are exponentially suppressed at early times  by the scale factors.} 
\be
\eta_0-\eta'\simeq \fr{1}{H_0}.\label{30}
\ee
If one furthermore takes $k$ to be superhorizon at $t_0$, the time integrals in \eq{28} can approximately be evaluated to yield
\be
P_{reg}(k,t_0)\simeq k-2c_1 e^2\sqrt{\fr{H}{H_0}}k+c_2\fr{e^2}{4\pi^2}\fr{H}{H_0}k,\hs{10}k\leq H_0,\label{31}
\ee
which is the main result of this paper. Note that when $k$ is superhorizon at $t_0$ in radiation, it must also be superhorizon at the end of inflation, i.e. $k\leq H_0$ implies that $k\leq H a(t_I)$. Compared to the flat space contribution, which is the first term in \eq{31}, the last term has the huge enlargement factor at late times (take, for example, the typically assumed inflationary energy scale $10^{16}$ GeV and a late time during radiation era with the energy scale $1$ MeV, where the corresponding Hubble parameters are $H\simeq10^{13}$ GeV and $H_0\simeq10^{-25}$ GeV). 

From the power spectrum \eq{31} one may calculate the field variance and the energy density of the {\it superhorizon modes} by restricting the range of the momentum integral to the superhorizon region $(0,H_0)$, which gives
\be
\left<B^2\right>_S\propto \r_S(t_0)\propto e^2HH_0^3.\label{32}
\ee 
Although $\r_S$ is small compared to the background energy density $6H_0^2M_p^2$, it is still much larger than the vacuum energy density of a free massless field that should be proportional to $H_0^4$. The amplitude of the magnetic field corresponding to \eq{32} is $B\propto e H_0\sqrt{HH_0}$, which involves the comparatively huge scale of inflation. The two point function in the position space, which is given by
\be
\vvc B_i(0)B_i(\vx)\vv =\fr{1}{2\pi^2r}\int_0^\infty dk\, k\,\sin(kr)\,P(k),
\ee
is also dominated by the superhorizon modes if $|\vx|=r\geq 1/H_0$, because  the subhorizon modes give oscillating contributions that are averaged out to zero. From \eq{31} one may deduce
\be
\vvc B_i(0)B_i(\vx)\vv \propto \,e^2\,\fr{H}{H_0}\,r^{-4},\hs{10}r\geq\fr{1}{H_0},
\ee
which expresses the superhorizon spectrum in the position space.  

Let us now use the full regularized power spectrum (not just the superhorizon approximation \eq{31})  to calculate the magnetic field variance in the position space, which includes all modes from UV to IR. As discussed in the previous section, the variance contains a distributional divergence that can be smoothed out by a window function as in \eq{19}. In the following we will take the window function \eq{21}, which allows us to carry out the momentum integrals analytically before taking the time integrals. Plugging  the regularized power spectrum, which is obtained  through \eq{25}, \eq{26a} and \eq{26}, in \eq{19} produces three distinct contributions to the variance. The cosmologically uninteresting tree level term gives a cutoff dependent result
\be
\vvc B_W^2\vv_I=\fr{3}{\pi^2\e^4}.\label{33}
\ee
After carrying out the momentum integral for the second term of \eq{25}, one may see that it's contribution has a finite $\e\to0$ limit giving
\be
\vvc B_W^2\vv_{II}=\fr{c_1e^2}{8\pi^2}H^2\int_{t_i}^{t_I}dt' \fr{a(t')}{(\eta_0-\eta')^3}.\label{34}
\ee
At late times when $H_0\ll H$, one may use the approximation \eq{30} and carry out the time integral to obtain
\be
\vvc B_W^2\vv_{II}\simeq\fr{c_1e^2}{8\pi^2}H_0^3\sqrt{HH_0}.\label{35}
\ee
We observe that because of the oscillatory nature of the term $\sin[2k(\eta_0-\eta')]$ in \eq{25}, the subhorizon modes do not alter the order of magnitude of the variance too much; nevertheless  they still suppress the corresponding negative superhorizon  contribution coming from \eq{31}. 

After carrying out the momentum integral coming from the third term of the power spectrum \eq{25} one gets
\bea
\vvc B_W^2\vv_{III}=&&\fr{e^2H^4}{16\pi^4} \int_{t_i}^{t_I}dt''\int_{t''}^{t_I}a(t')\,a(t'')\nn\\
&&\int_0^\infty l\,dl\, \int_{-1}^{1}du\,\fr{(1-u^2)}{(1+l^2+2lu)^{3/2}}
\left[\fr{\e^2-\a^2}{(\e^2+\a^2)^2}-\fr{\e^2-\b^2}{(\e^2+\b^2)^2}\right],\label{36}
\eea
where 
\bea
&&\a=(\eta'-\eta'')\left(1+l+\sqrt{1+l^2+2lu}\right),\nn\\
&&\b=2\eta_0-\eta'-\eta''+(\eta'-\eta'')\left(l+\sqrt{1+l^2+2lu}\right).\label{37}
\eea
We estimate this contribution as follows: First we note that the integrand is suppressed by $1/l^2$ for $l\gg1$, therefore the $l$-integral gets its largest contribution near $l=O(1)$. Similarly, the $u$-integral also gives an order one contribution. Therefore, to estimate \eq{36} it is enough to determine the magnitude of the integrand near $l=O(1)$ and $u=O(1)$. Because of the scale factors, the time integrals become largest  near the end of inflation since earlier times are exponentially suppressed. In terms of the dimensionless variables $Ht'$ and $Ht''$, the upper limits of the time integrals become $1/2$ by \eq{23}, and the scale factors can be approximated as $a(t')\simeq a(t'')\simeq a_I$. Finally, one may check that
\be
0\leq \eta'-\eta''\leq\fr{1}{\sqrt{HH_0}},\label{38}
\ee
and again $H_0\ll H$ implies $\eta_0-\eta'\simeq\eta_0-\eta''\simeq1/H_0$. As a result, in estimating \eq{36} one may use
\be
\b\simeq\fr{2}{H_0},\hs{10}0\leq \a<O\left(\fr{1}{\sqrt{HH_0}}\right).\label{39}
\ee
For the second term in the square bracket in \eq{36}, $\e\to0$ limit can safely be taken and the corresponding contribution to the variance has the order of magnitude $e^2\,H_0^3\sqrt{HH_0}$, which is comparable to \eq{35}. 

On the other hand, the first term in the square brackets in \eq{36} becomes singular in the $\e\to0$ limit since the double time integral diverges on the line $\eta'=\eta''$ giving $\a=0$. Therefore, the magnitude of this term depends on the  cutoff scale $\e$. There is no general principle that  fixes the value of $\e$, usually it is chosen according to the physical observable of interest. For example, if $t_0$ denotes today and one is dealing with the intergalactic astrophysical observations, it would be natural to set $\e$ to be the size of a typical galaxy, call $r_G$. In that case $r_G\gg\a$ and \eq{36} can be estimated as
\be
\vvc B_W^2\vv_{III}\propto \fr{e^2H_0\,H}{r_G^2},\label{40}
\ee
which is much larger than both \eq{32} and \eq{35}. Summarizing, we find that the one loop Higgs correction to the photon two point function yields a large scale magnetic field background \eq{40} corresponding to the averaging scale $r_G$. Using $H_0\simeq 10^{-33}eV$, and taking the reasonable value $H\simeq10^{16}GeV$ and the averaging scale $r_G=1kpc\simeq 3\times10^{21}cm$, \eq{40} gives $B_W\propto e\, 10^{-28}\hs{2}Gauss$, where one can utilize the conversion factors $1/cm\simeq 2\times10^{-5}eV$ and $Gauss\simeq 2\times10^{-2}eV^2$. Here, the value of the coupling constant $e$ at the scale of inflation must be determined by the renormalization group flow in the $SU(2)\times U(1)$ theory. Alternatively, \eq{39} suggests the cutoff scale $\sqrt{HH_0}$ since $\a$ is the variable yielding the divergence in \eq{36}, and this gives $B_W\propto e\, 10^{-6}\hs{2}Gauss$. Obviously, at earlier times  the magnetic field becomes larger. 

\section{Conclusions}

In this paper we calculate the first order Higgs loop correction to the late time magnetic field power spectrum in inflationary cosmology. Using the in-in perturbation theory, this  loop effect can be expressed in terms of the photon and the scalar field mode functions. We observe that the loop time integrals can be divided into non-overlapping intervals corresponding to the distinct epochs and we argue that the dominant contribution comes from the period of inflation. As a result, the power spectrum at a late time can be approximately determined from the well known inflationary scalar mode functions. Since the photons are conformally coupled to gravity, the photon mode function is identical to the flat space counterpart at all times during cosmological history, which helps us to pin down the late time behavior of the power spectrum. 

In our calculation, we utilize the adiabatic regularization to cure the loop infinities. The massless de Sitter  scalar mode function running in the loop can be decomposed into the adiabatic order zero flat space and adiabatic order one expanding pieces. With this decomposition one can easily apply the adiabatic regularization by throwing out the divergent integrands that do not have enough adiabaticity. We find that at superhorizon scales at some later time after inflation, the regularized power spectrum becomes much larger than the tree level result by a factor $H/H_0$, where $H$ and $H_0$ are the Hubble parameters of inflation and the time of interest. In the appendix we apply dimensional regularization of the loop integrals and show that the results agree with the adiabatic method. Therefore, the calculated superhorizon behavior of the electromagnetic field induced by the Higgs scalar must be a robust physical effect as long as the simple model \eq{1} is valid. 

Although the spectrum is suppressed at IR compared to the scale free behavior (while our spectrum is proportional to $k$, a scale free one is given by $1/k^3$), the presence of the huge superhorizon enlargement factor is encouraging (note that the scale free power spectrum of a massless scalar field is given by $H^2/(2k^3)$ and the superhorizon energy density corresponding to the kinetic term $(\del_i\f)^2$ equals $H^2H_0^2$, which is larger compared to \eq{32}). Therefore, it would be interesting to examine whether this loop correction is large enough to produce any cosmologically interesting effects. For example, one may check if the curious large scale micro-Gauss intergalactic magnetic fields (see e.g. \cite{g1,g2,g3}) can be accounted for by this mechanism. Without doubt, looking for potential imprints on the CMB radiation is another alternative. Apparently, two possibilities are worth to study; an instant imprint at the time of decoupling and a cumulative impact on the propagating photons from the time of decoupling to the present time via photon-photon interactions. One must note, however, that extra work is needed to pin down any residual effect; although promising, the superhorizon enlargement factor found in this paper does not guarantee any measurable contribution. The corresponding superhorizon energy density is generically small so no significant effect should be expected involving gravitational physics. 

Another interesting question is to see how symmetry breaking possibly modifies the present loop effect. Recall that in our calculation we have neglected the Higgs vacuum expectation value and studied the unbroken theory because the presumed scale of inflation is usually much larger than the  symmetry breaking scale. Clearly, the theory changes significantly following symmetry breaking some time after inflation. Nevertheless, the main argument of this paper is that the late time Higgs loop correction can be divided into the sum of different epochs and the period of inflation should give the largest contribution. Moreover, as discussed below \eq{25}, the inflationary contribution is transferred to future by the photon mode functions, which are not modified by symmetry breaking since the photons are always massless and their coupling to gravity is conformal. Therefore, one expects that any physics after inflation must yield only subleading corrections and it would be interesting to verify this expectation for symmetry breaking.  
 
\appendix*

\section{Dimensional Regularization}

In this appendix we carry out the dimensional regularization of the power spectrum \eq{25}. First, let us note that the massless scalar mode equation in $d$-spatial dimensions becomes  
\be
\ddot{\f}_k^{(d)}+d\,\fr{\dot{a}}{a}\,\dot{\f}_k^{(d)}+\fr{k^2}{a^2}\f_k^{(d)}=0,\hs{10}\f_k^{(d)}\dot{\f}_k^{(d)*}-\f_k^{(d)*}\dot{\f}_k^{(d)}=\fr{i}{a^d}.\label{a1}
\ee
In de Sitter space, \eq{a1} can be solved for the Bunch-Davies vacuum as 
\be
\f_k^{(d)}=\fr{1}{a^{d/2}}\sqrt{\fr{\pi}{4H}}\exp\left(i\pi d/2\right)H_{d/2}^{(1)}\left(\fr{k}{aH}\right).\label{a2} 
\ee
The scalar-photon coupling constant acquires a nonzero mass dimension and we define
\be\label{a3}
e_\m=\m^{\d/2}e,\hs{10}\d\equiv3-d,
\ee
where $\m$ is an arbitrary mass parameter. 

In $d$-spatial dimensions the second term in \eq{25} generalizes to
\be\label{a4}
II\equiv e_\m^2\vvc \f^\dagger\f\vv=\fr{e_\m^2}{(2\pi)^d}\int\, d^d k\,|\f_k^{(d)}|^2.
\ee
Using \eq{a2} and the scaling $k=aH\hat{k}$, \eq{a4} becomes
\be
II=\fr{e^2H^2}{32\pi^2}\left(\fr{2\pi\m}{H}\right)^\d\int\, d^{3-\d}\hk\,|H_{(3-\d)/2}^{(1)}(\hk)|^2.
\ee
For the dimensional regularization to work out, the integral must obey 
\be\label{a5}
\lim_{\d\to0}\,\int\, d^{3-\d}\hk\,|H_{(3-\d)/2}^{(1)}(\hk)|^2=\fr{F_0}{\d}+F_1,
\ee
where $F_0$ and $F_1$ are finite numerical constants. Indeed, the Hankel function has the following integral representation
\be\label{aint}
H_n^{(1)}(\hk)=\fr{\C[\fr12-n](\fr12 \hk)^n}{\pi^{\fr32}i}\oint_{1+i\infty}^{(1^+)}e^{i\hk v}(v^2-1)^{n-\fr12}dv,
\ee
where the simple loop contour starts at $v=1+i\infty$ in the complex $v$-plane, circles $v=1$ once in the counter-clockwise direction and returns to $v=1+i\infty$. This integral representation is valid for any non-half-integer $n$ and when $\arg(\hk)<\pi/2$. Crucially, the contour can be chosen so that $v$ has always a positive imaginary piece and thus $\exp(i\hk v)$ gives an exponential dumping. Let us use \eq{aint} for one of the Hankel functions in \eq{a5}. One may see that the integrals are well defined as $\d\to0$; specifically $d^{3-\d}\hk$ integral converges as $\hk\to\infty$ because of the exponential dumping coming from $\exp(i\hk v)$. Therefore, \eq{aint} allows one to hide the loop divergence in the gamma-function and \eq{a5} can be written as 
\be
\lim_{\d\to0}\,\int\, d^{3-\d}\hk\,|H_{(3-\d)/2}^{(1)}(\hk)|^2=\lim_{\d\to0}\,\C\left[\fr{\d-2}{2}\right]G(\d),
\ee
where $G(\d)$ is an analytic function of $\d$. This proves \eq{a5} where the constants $F_0$ and $F_1$ can be read from the expansions of $G(\d)$ and the gamma-function $\C[(\d-2)/2]$ in the limit $\d\to0$. As a result, \eq{a4} becomes
\be
II=\fr{e^2H^2}{32\pi^2}\,\lim_{\d\to0}\,\left(1+\d\ln(2\pi\m/H)\right)\left(F_0/\d+F_1\right).
\ee 
The singular $1/\d$ piece can be canceled out by a mass counterterm and the remaining terms imply
\be
e^2\vvc \f^\dagger\f\vv_{reg}=e^2c_1H^2,\label{a26a}
\ee
where $c_1=[F_0\ln(2\pi\m/H)+F_1]/(32\pi^2)$; this qualitatively agrees with the result \eq{26a} obtained by adiabatic regularization.  

For the third term in \eq{25}, the relevant loop integral in $d$-spatial dimensions becomes
\be
III\equiv e_\m^2F(k,t',t'')=\fr{e_\m^2}{(2\pi)^d}\int 
d^d\tk\,\,\tk_i\,\tk_j\left(\d_{ij}-\fr{k_ik_j}{k^2}\right)\,\f_{\tk}^{(d)}(t')\,\f_{\tk}^{(d)*}(t'')\,\f_{|\vec{k}+\vec{\tk}|}^{(d)}(t')\,\f_{|\vec{k}+\vec{\tk}|}^{(d)*}(t'').\\
\ee
Using \eq{a2}, one may obtain  
\bea
III=&&\fr{e_\m^2}{(2\pi)^d}\fr{\pi^2}{16H^2a(t')^da(t'')^d}\int 
d^d\tk\,\,\tk_i\,\tk_j\left(\d_{ij}-\fr{k_ik_j}{k^2}\right)\,\nn\\
&&H_{d/2}^{(1)}\left(\fr{\tk}{a(t')H}\right)H_{d/2}^{(1)*}\left(\fr{\tk}{a(t'')H}\right)H_{d/2}^{(1)}\left(\fr{|\vec{k}+\vec{\tk}|}{a(t')H}\right)H_{d/2}^{(1)*}\left(\fr{|\vec{k}+\vec{\tk}|}{a(t'')H}\right).\label{a12}
\eea
To extract the leading order behavior as $k\to0$, which is the most relevant one for superhorizon physics, we first define $\vec{\tk}=k \vec{l}$. Then, using the small argument expansion\footnote{Since the dimensional regularization ensures that \eq{a12} is well defined, it does not matter whether one first apply the expansion of the Hankel functions and then calculate the momentum integral, or the other way around.} of the Hankel function
\be\label{a13}
\lim_{z\to0}H_n^{(1)}(z)=-\fr{i}{\pi}2^n\C[n]\,z^{-n}-\fr{i}{\pi}2^{-2+n}\C[-1+n]\,z^{-n+2}+...
\ee
in \eq{a12}, one may see that the leading order contribution, which comes from the first expansion term in \eq{a13}, becomes
\be
III_1=\left(\fr{2}{\pi}\right)^d\fr{e^2\O_{d-2}}{16\pi^2}\C[d/2]^4\left(\fr{H^4}{k}\right)\left(\fr{\m k}{H^2}\right)^\d\int_0^\infty l\, dl\, \int_{-1}^{1}du\fr{(1-u^2)^{d/2-1}}{(1+l^2+2lu)^{d/2}},\label{ar1}
\ee
where $\O_{d-2}$ is the area of unit $(d-2)$-sphere, $u=\cos(\th)$ and $\th$ is the angle between $\vec{k}$ and $\vec{l}$. In getting that result we have used spherical coordinates in momentum space so that 
\be
\int 
d^d\tk\,\,\tk_i\,\tk_j\left(\d_{ij}-\fr{k_ik_j}{k^2}\right)=\int \tk^{d+1}\sin^{d-2}(\th)(1-\cos^2\th)\,d\tk \,d\th \,d\O_{d-2}.
\ee
One may observe that \eq{ar1} is finite as $\d\to0$, which gives 
\be\label{a16}
III_1=c_2\fr{e^2H^4}{16\pi^2k},
\ee
where $c_2=\int_0^\infty l\,dl\, \int_{-1}^{1}du(1-u^2)/(1+l^2+2lu)^{3/2}$. This result is exactly the same with \eq{27}, which shows the equivalence of the dimensional and adiabatic regularizations for this specific loop correction.  

It is instructive to determine the next to leading order contribution by keeping the second term in \eq{a13} while expanding the Hankel functions in \eq{a12} (in the adiabatic regularization, this corresponds to the expansion of  the exponential term in \eq{26}) A straightforward calculation gives
\bea
III_2=\left(\fr{2}{\pi}\right)^d&&\fr{e^2\O_{d-2}}{64\pi^2}\C[d/2]^3\C[d/2-1](H^2k)\left(\fr{\m k}{H^2}\right)^\d\nn\\
&&\left[\fr{a(t')^2+a(t'')^2}{a(t')^2a(t'')^2}\right]\,\int_0^\infty l\, dl\, \int_{-1}^{1}du\,(1-u^2)^{\fr{d-1}{2}}\,\fr{(1+2l^2+2lu)}{(1+l^2+2lu)^{d/2}}.\label{a17}
\eea
By comparing to \eq{ar1}, one sees that \eq{a17} is smaller by the factor $k^2/(a^2H^2)$ (note that superhorizon modes obey $k/aH\ll1$), which justifies the small argument expansion of \eq{a12}. As $\d\to0$ the integral in \eq{a17} must approach $E_0/\d+E_1$ for some finite numerical constants $E_0$ and $E_1$. As usual, the singular term in \eq{a17} can be canceled out by a coupling constant renormalization and the finite parts primarily yield a factor $e^2H^2k(E_1+E_0\ln(\m k/H^2))$.    

To sum up, we show that in our case the adiabatic and dimensional regularizations agree with each other. Specifically, the regularized expressions \eq{27} and \eq{a16} turn out to be identical, and the corrections \eq{26a} and \eq{a26a} are consistent with each other. In principle, \eq{26a} and \eq{a26a} can also be made identical by choosing the renormalization scale $\m$ in \eq{a26a} appropriately (of course, in doing that one must also take into account the renormalization conditions). The rest of the calculation following \eq{26a} and \eq{27}, which solely involves time integrals, is identical in both regularization schemes.  

\begin{acknowledgments}
This work, which has been done at C.\.{I}.K. Ferizli, Sakarya, Turkey without any possibility of using references, is dedicated to my friends at rooms C-1 and E-10 who made my stay bearable at hell for 440 days between 7.10.2016 and 20.12.2017. I am also indebted to the colleagues who show support in these difficult times. I thank to the anonymous referee for various points raised, which improved the clarity of the paper.  
\end{acknowledgments}

\end{document}